\documentclass[journal,comsoc]{IEEEtran}

\usepackage{amsmath}
\usepackage{amssymb}
\usepackage{amsfonts}
\usepackage{color}
\usepackage{graphicx}
\usepackage{algorithm}
\usepackage{algorithmic}
\usepackage{enumerate}
\usepackage{cite}
\usepackage{multirow}
\usepackage{lipsum,mwe,cuted}
\usepackage{diagbox}
\usepackage{graphicx}
\usepackage{float}
\usepackage{subfig}

\allowdisplaybreaks
\begin{document}
\title{A Message Passing Detection based Affine Frequency Division Multiplexing Communication System}
\author{\IEEEauthorblockN{Lifan Wu\IEEEauthorrefmark{1}, Shan Luo\IEEEauthorrefmark{1}, Dongxiao Song\IEEEauthorrefmark{1}, Fan Yang\IEEEauthorrefmark{1} and Rongping Lin\IEEEauthorrefmark{2}\\
\IEEEauthorblockA{\IEEEauthorrefmark{1}School of Aeronautics and Astronautics,\\
\IEEEauthorrefmark{2}School of Information and Communication Engineering,\\
University of Electronic Science and Technology of China, Chengdu, China,\\
Email: lifan\_w@163.com, luoshan@uestc.edu.cn, songdongxiao0813@foxmail.com, yangfan000209@163.com, linrp@uestc.edu.cn}}
\vspace*{-10mm}
\vspace*{-1mm}\thanks{This work was supported by the National Natural Science Foundation of China (NSFC) (grant number 61871097) and the Sichuan Science and Technology Program (grant number 2023YFG0298).}}

\maketitle
\begin{abstract}
  The next generation of wireless communication technology is anticipated to address the communication reliability challenges encountered in high-speed mobile communication scenarios. An orthogonal time frequency space (OTFS) system has been introduced as a solution that effectively mitigates these issues. However, OTFS is associated with relatively high pilot overhead and multiuser multiplexing overhead. In response to these concerns within the OTFS framework, a novel modulation technology known as affine frequency division multiplexing (AFDM) which is based on the discrete affine Fourier transform has emerged. AFDM effectively resolves the challenges by achieving full diversity through parameter adjustments aligned with the channel's delay-Doppler (DD) profile. Consequently, AFDM is capable of achieving performance levels comparable to OTFS. As the research on AFDM detection is currently limited, we present a message passing (MP) algorithm that is characterized by both low-complexity and efficiency. The algorithm exploits the inherent sparsity of the channel to efficiently handle joint interference cancellation and detection. Based on simulation results, the MP detection outperforms minimum mean square error (MMSE) and maximal ratio combining (MRC) detection.
\end{abstract}



\section{Introduction}	
The upcoming 6G mobile communication technology is poised to address the reliability concerns associated with communication in scenarios characterized by high mobility. Examples of such scenarios include high-speed rail transport, unmanned aerial vehicles (UAVs), and vehicle-to-vehicle communications\cite{white2}. Nevertheless, the presence of substantial Doppler shifts in high-speed mobile environments often leads to multipath fading. The significant Doppler effect, particularly on orthogonal frequency division multiplexing (OFDM), brought about by high mobility or time-varying channels, can introduce considerable distortion and ultimately degrade the performance of demodulation\cite{OFDM}.
\par In order to mitigate the effects of Doppler shift on communication, various multicarrier modulation techniques have been proposed. One of these techniques is orthogonal chirp division multiplexing (OCDM)\cite{OCDM}. OCDM employs the discrete Fresnel transform, which can be achieved through the discrete affine Fourier transform (DAFT) with specific parameters. Due to its higher diversity order, OCDM outperforms OFDM in time-dispersive channels. Nonetheless, the diversity order of OCDM hinges on the delay-Doppler (DD) profile of the channel, thereby constraining its capacity to achieve full diversity in generic linear time-varying (LTV) channels. Another notable two-dimensional (2D) modulation technique, known as orthogonal time frequency space (OTFS)\cite{OTFS}, was introduced by Hadani. This approach facilitates the conversion of information from the DD coordinate system to the more conventional time-frequency domain. OTFS is well-suited for conventional modulation schemes like OFDM, CDMA, and TDMA. By incorporating full diversity in both time and frequency domains, OTFS and its equalization techniques alleviate the fading and time-varying effects encountered by modulated signals such as OFDM.
\par Researchers have shown keen interest in the OTFS and explored various aspects of its performance. Some of the topics studied include coded OTFS\cite{code1,code2}, schemes for detection\cite{detection2,detection3}, peak-to-average power ratio (PAPR)\cite{PAPR1,PAPR2,PAPR3} and channel estimation\cite{channel1,channel2}, etc. The results indicate that OTFS is an effective modulation scheme in high mobility scenarios with large Doppler shifts\cite{anms}.
\par However, OTFS also has drawbacks, as it uses 2D transform that increases pilot overhead and multiuser multiplexing overhead\cite{epce}. To overcome these limitations, Ali Bemani et al. proposed affine frequency division multiplexing (AFDM) in 2021 as an alternative. The AFDM is a novel modulation scheme based on the traditional OFDM but with significant improvements to enhance the performance of wireless communication systems. The AFDM achieves full diversity by adapting parameters based on the channel's DD profile, leading to performance comparable to the OTFS\cite{AFDM1,AFDM2}. In the AFDM, symbols are efficiently multiplexed onto a collection of orthogonal chirps that generate a full and sparse DD representation of the channel in the DAFT domain by adapting to the doubly dispersive channel characteristics\cite{amab,cdf}. Research results presented in\cite{AFDM1,AFDM2}, \cite{AFDM2} demonstrate that the AFDM exhibits exceptional performance on par with OTFS while offering the distinct advantage of reduced channel estimation overhead.
\par Currently, research on AFDM is rapidly expanding, and the published work primarily focuses on the following areas. Pilot-aided channel estimation has been proposed in\cite{AFDMce}, aiming to accurately estimate the channel characteristics in AFDM systems. Low complexity equalization techniques have been proposed in\cite{AFDMeq}, addressing the challenges of efficient equalization in the AFDM systems. An integrated sensing and communications approach based on the AFDM has been proposed in\cite{AFDMsc}, exploring the potential of the AFDM in combined sensing and communication applications. However, it is noted that research on the detection method for the AFDM is relatively limited. Further exploration and development in this area could contribute to advancing the understanding and practical implementation of the AFDM in various communication scenarios.
\par Hence, our primary focus in this paper is on the detection method for AFDM. We introduce the message passing (MP) algorithm, a method characterized by its low complexity and high efficiency. This algorithm combines interference cancellation (IC) and detection, capitalizing on the inherent sparsity of the channel. To achieve further reduction in complexity, a sparse factor graph is employed, coupled with the utilization of Gaussian approximation for interference terms. This strategy draws inspiration from a comparable technique employed in \cite{lcd}, which offers the advantage of direct applicability to large-scale MIMO systems without factoring in channel sparsity. Within the framework of the MP algorithm, the potential to mitigate inter-carrier interference (ICI) and inter-symbol interference (ISI) through precise phase shifting becomes apparent. Conversely, the challenge of inter-Doppler interference (IDI) can be effectively addressed by utilizing the MP algorithm to prioritize the most significant interference factors. As a result, a broad spectrum of channel Doppler spreads can be effectively mitigated by the proposed MP algorithm\cite{mpa}. Extensive simulation exercises have shown that the performance of AFDM when utilizing the proposed MP algorithm aligns closely with that of OTFS. This observation underscores AFDM's capability as a robust modulation scheme for scenarios involving high mobility, characterized by considerable Doppler shifts.
\par The following sections of this paper are organized as follows. The concepts of affine Fourier transform (AFT) and DAFT and the complete transmission process of the AFDM system are introduced in Section II. Section III provides a comprehensive explanation of the MP algorithm. In Section IV, the simulation results are analyzed, highlighting the performance of AFDM. Ultimately, Section V concludes this paper.
\section{RELATED WORK}
This section provides a comprehensive review of two essential concepts: AFT and DAFT. These concepts lay the foundation for AFDM. Additionally, we introduce relevant notions and pivotal elements of the AFDM.
\subsection{Affine Fourier Transform}
\subsubsection{Continuous Affine Fourier Transform}
The AFT, also called the linear canonical transform\cite{lct}, is a linear integral transform characterized by four parameters and serves as the basis for AFDM. The AFT is mathematically defined as follows
\begin{equation}
  \label{S}
  S_{a,b,c,d}\left( u \right) =\begin{cases}
    \int_{-\infty}^{+\infty}{s\left( t \right)}K_{a,b,c,d}\left( t,u \right) \mathrm{d}t&		,b\ne 0\\
    s\left( du \right) \frac{e^{-jcdu^2/2}}{\sqrt{a}}&		,b=0\\
  \end{cases}
\end{equation}
where the parameters $(a,b,c,d)$ constitutes a matrix
$\mathbf{M}=\left[ \begin{matrix}
	a&		b\\
	c&		d\\
\end{matrix} \right] $
, and $\left |\mathbf{M} \right |$ is $1$, i.e $ad-bc=1$, and the core of this transformation is
\begin{equation}
  K_{a,b,c,d}\left( t,u \right) =\frac{1}{\sqrt{2\pi \left| b \right|}}e^{-j\left( \frac{a}{2b}u^2+\frac{1}{b}ut+\frac{d}{2b}t^2 \right)}
\end{equation}
By using the parameter 
$\mathbf{M}^{-1}=\left[ \begin{matrix}
	d&		-b\\
	-c&		a\\
\end{matrix} \right] $
, the AFT can be represented as the inverse AFT (IAFT) as follows
\begin{equation}
  s\left( t \right) =\int_{-\infty}^{+\infty}{S_{a,b,c,d}\left( u \right) K_{a,b,c,d}^{*}\left( t,u \right) \mathrm{d}u}
\end{equation}
\par Numerous widely recognized mathematical transformations stem from the AFT. For instance, when the parameter is set to $(0,1/2\pi,-2\pi,0)$, the AFT corresponds to the standard Fourier transform. Similarly, with a parameter of $(0,j(1/2\pi),j2\pi,0)$, the AFT becomes the Laplace transform. When the parameter is defined as $(\cos \theta,(1/2\pi) \sin \theta,-2\pi \sin \theta,\cos \theta)$, the AFT embodies the $\theta$ Fractional Fourier transform. The additional degrees of freedom present in the AFT introduce versatility and have discovered applications across diverse domains.
\par The versatility of the AFT allows researchers and engineers to adapt it to specific use cases and tailor it to address various challenges in signal processing and communication systems. This flexibility has contributed to the development of novel modulation techniques like the AFDM, which exploits AFT's properties to overcome the challenges posed by high-mobility scenarios with significant Doppler shifts.
\vspace{1mm}
\subsubsection{Discrete Affine Fourier Transform}
The DAFT serves as the discrete equivalent of the AFT, facilitating the calculation of continuous transformations in tasks like spectral analysis or discrete data signal processing. In the spectral analysis scenario, the continuous function is sampled to provide the input for the discrete transformation. On the other hand, in the case of discrete data processing, the input consists solely of a discrete sequence.
\par We designate the input signal as $s(t)$ and its corresponding AFT as $S_{a,b,c,d}(u)$. These are sampled at intervals $\Delta t$ and $\Delta u$, respectively, to enable discrete transformation. This discrete variant of AFT, known as the DAFT, proves valuable in a variety of practical applications, accommodating scenarios where input data is either continuous and sampled or entirely discrete. By sampling $s(t)$ at intervals of $\Delta t$ and $S_{a,b,c,d}(u)$ at intervals $\Delta u$, we obtain
\begin{equation}
  \label{sS}
  \begin{aligned}
  s[n]&=s\left( n\Delta t \right) \\
  S[m]=&S_{a,b,c,d}\left( m\Delta u \right)
  \end{aligned}
\end{equation}
\begin{figure*}[h]
  \centering
  \includegraphics[width=\linewidth]{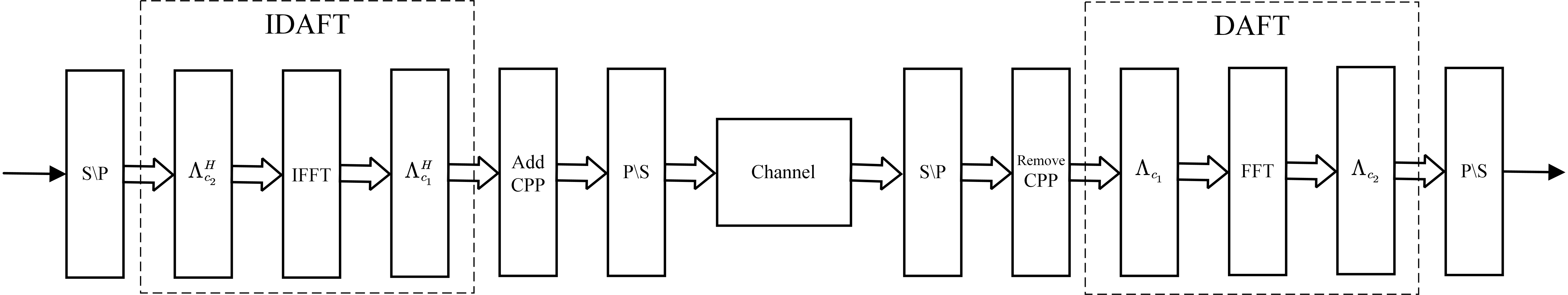}
  \caption{Block diagram of the AFDM system}
\end{figure*}
where $n$ and $m$ are integers, and $0\le n \le N-1$, $0\le m \le M-1$. According to $(\ref{sS})$, $(\ref{S})$ can be converted into
\begin{equation}
  \label{Sm}
  S[m]=\frac{\Delta t}{\sqrt{2\pi \left| b \right|}} \cdot e^{-j\left( \frac{a}{2b}m^2\Delta u^2 \right)}\sum_{n=0}^{N-1}{e^{-j\left( \frac{1}{b}mn\Delta u\Delta t+\frac{d}{2b}n^2\Delta t^2 \right)}s[n]}
\end{equation}
We can write $(\ref{Sm})$ in the form of a matrix as follows
\begin{equation}
  \label{Sm1}
  S[m]=\sum_{n=0}^{N-1}{F_{a,b,c,d}\left[ m,n \right] s[n]}
\end{equation}
where
$$
F_{a,b,c,d}\left[ m,n \right] =\frac{1}{\sqrt{2\pi \left| b \right|}}\cdot \Delta t\cdot e^{-j\left( \frac{a}{2b}m^2\Delta u^2+\frac{1}{b}mn\Delta u\Delta t+\frac{d}{2b}n^2\Delta t^2 \right)}
$$
\par In order to make $(\ref{Sm1})$ reversible, the following conditions should hold
\begin{equation}
  \label{dtdu}
  \Delta t\Delta u=\frac{2\pi \left| b \right|}{M}
\end{equation}
Therefore, we can express the first type of the DAFT as follows
\begin{equation}
  \label{Sm2}
  S[m]=\frac{1}{\sqrt{M}}e^{-j\frac{a}{2b}m^2\Delta u^2}\sum_{n=0}^{N-1}{e^{-j\left( \text{sgn}(b) \frac{2\pi}{M}mn+\frac{d}{2b}n^2\Delta t^2 \right)}s[n]}
\end{equation}
\par The second type of DAFT is obtained by defining $c_1=\frac{d}{4\pi b}\Delta t^2$ and $c_2=\frac{a}{4\pi b}\Delta u^2$, therefore, $S[m]$ in $(\ref{Sm1})$ is written as $S[m]=\sum_{n=0}^{N-1}{F_{c_1,c_2}\left[ n,m \right] s[n]}$, where
\begin{equation}
  \label{Fc1c2}
  F_{c_1,c_2}\left[ n,m \right] \triangleq \frac{1}{\sqrt{M}}e^{-j2\pi \left( c_2m^2+\frac{\text{sgn} \left( b \right)}{M}mn+c_1n^2 \right)}
\end{equation}
\par The condition given by $(\ref{dtdu})$ can be expressed as $c_1c_2=\frac{ad}{4M^2}$. This condition remains applicable, as there are no restrictions on the values of $a$ and $d$, allowing for the consideration of any real numbers. Hence, provided that the condition $ad-bc=1$ holds for $b$ and $c$, there are no constraints on the values of $c_1$ and $c_2$, allowing for the consideration of any real number. Simplifying $(\ref{Sm2})$ can be achieved by assuming $\text{sgn}(b)=1$. Therefore, the DAFT can be written as follows
\begin{equation}
  \label{Sm3}
  S[m]=\frac{1}{\sqrt{M}}e^{-j2\pi c_2m^2}\sum_{n=0}^{N-1}{e^{-j2\pi \left( \frac{1}{M}mn+c_1n^2 \right)}s[n]}
\end{equation}
where $M\ge N$ and the inverse DAFT (IDAFT) can be written as follows
\begin{equation}
  \label{sn}
  s[n]=\frac{1}{\sqrt{M}}e^{j2\pi c_1n^2}\sum_{n=0}^{M-1}{e^{j2\pi \left( \frac{1}{M}mn+c_2m^2 \right)}S[m]}
\end{equation}
From $(\ref{Sm3})$ and $(\ref{sn})$, it can be seen that the periodicity is as follows
\begin{equation}
  \label{Sm+k}
  S[m+kM]=e^{-j2\pi c_2\left( k^2M^2+2kMm \right)}S[m]
\end{equation}
\begin{equation}
  \label{sn+k}
  s[n+kN]=e^{j2\pi c_1\left( k^2N^2+2kNn \right)}s[n]
\end{equation}
\par Of the two properties mentioned above, only $(\ref{sn+k})$ is significant, as it has an impact on the prefix types that should be added to DAFT-based multicarrier symbols. When $M=N$, the inverse transform remains identical to the forward transform, with the parameters being $-c_1$ and $-c_2$, alongside conjugating the Fourier transform term.
\par Assuming there are time domain signals $$\mathbf{s}=\left( s_0,s_1,...,s_{N-1} \right)^T$$ and its corresponding DAFT signals $$\mathbf{S}=\left( S_0,S_1,...,S_{N-1} \right)^T$$, we can write the DAFT in matrix form as follows
\begin{equation}
  \label{S=As}
  \mathbf{S}=\mathbf{As}
\end{equation}
where $\mathbf{A}=\mathbf{\Lambda}_{c_2}\mathbf{F}\mathbf{\Lambda}_{c_1}$, $\mathbf{\Lambda}_c$ is defined as
\begin{equation}
  \setcounter{equation}{15}
  \mathbf{\Lambda}_c=\mathrm{diag}\left( e^{-j2\pi cn^2},n=0,1,\dots,N-1 \right)
\end{equation}
and $\mathbf{F}$ is the DFT matrix that having entries $e^{-j2\pi mn/N}/\sqrt{N}$. Given that both matrix $\mathbf{\Lambda}_c$ and matrix $\mathbf{F}$ are unitary, it follows that matrix $\mathbf{A}$ is also unitary. The inverse of the matrix $\mathbf{A}$ can be represented as $\mathbf{A}^{-1}=\mathbf{A}^H=\mathbf{\Lambda}_{c_1}^{H}\mathbf{F}^H\mathbf{\Lambda }_{c_2}^{H}$.

\subsection{Affine Frequency Division Multiplexing}
The AFDM is a multicarrier modulation concept that employs DAFT. At the transmitting end, the message signal is subjected to modulation through IDAFT, converting it into a time-domain signal. Upon reception, the signal is demodulated using DAFT to recover the original message signal. This process is shown in Fig. 1.

\subsubsection{Modulation}
Suppose that the message signal $\mathbf{x} \in \mathbb{A}^{N\times 1}$ is a vector in the DAFT domain. Here, $\mathbb{A}$ represents the alphabet of quadrature amplitude modulation (QAM) symbols, and its constituents are numerical entities of the structure $R+Ij$, where $R$ and $I$ denote integers. The message signal $x[n]$ transforms into a sending signal $s[n]$ through IDAFT, as follows
\begin{equation}
  \label{s[n]}
  s\left[ n \right] =\frac{1}{\sqrt{N}}\sum_{m=0}^{N-1}{e^{j2\pi \left( c_1n^2+\frac{nm}{N}+c_2m^2 \right)}\cdot x\left[ m \right]} 
\end{equation}
where $n=0,\dots ,N-1$. We can write $(\ref{s[n]})$ in the form of a matrix as $\mathbf{s}=\mathbf{A}^H\mathbf{x}=\mathbf{\Lambda} _{c_1}^{H}\mathbf{F}^H\mathbf{\Lambda} _{c_2}^{H}\mathbf{x}$.
\par In order to counteract the impacts of multipath propagation and create the illusion of a periodic domain in the channel, we prepend a {\em chirp-periodic} prefix (CPP) the time-domain transmission signal due to different signal periodicity, rather than an OFDM periodic prefix (CP). The length of the CPP which is denoted as $L$, is selected as an integer greater than or equal to the maximum channel delay spread of the sample. By utilizing the periodicity defined in $(\ref{sn+k})$, we can write the prefix as follows
\begin{equation}
  s\left[ n \right]=s\left[ N+n \right] e^{-j2\pi c_1\left( N^2+2Nn \right)},   n=-L,\cdots ,-1
\end{equation}
When $2Nc_1$ takes on an integer value and $N$ is an even number, the CPP essentially becomes a cyclic prefix (CP).
\subsubsection{Channel}
After channel transmission, serial to parallel, removing the CPP, the received signal in the DAFT domain is
\begin{equation}
  \label{r[n]}
  r\left[ n \right] =\sum_{i=1}^P{h_ie^{-j2\pi f_in}s\left[ n-l_i \right] +w\left[ n \right]}
\end{equation}
where $P \geq 1$ denotes the number of paths, $h_i$ stands for the complex gain associated with the $i$-th path, $l_i$ corresponds to the delay of the $i$-th path normalized with respect to the sample period, and $f_i$ represents the Doppler shift of the same path. Furthermore, $w[n]$ represents the additive Gaussian noise.
\par Assuming that $\nu_i$ denotes the Doppler shift, which has been normalized in relation to the subcarrier interval of the $i$-th path. It is worth noting that, in this paper, we treat $\nu_i$ as an integer. This leads to the definition $\nu_i \triangleq Nf_i \in \left[-\nu_{\max}, \nu_{\max}\right]$, where $\nu_{\max}$ signifies the maximum Doppler shift within the LTV channel. Additionally, we make the assumption that the greatest delay present in the channel, denoted as $l_{\max}$, satisfies the condition $l_{\max} < N$. Furthermore, it is ensured that the length of the CPP is greater than or equal to $l_{\max}$.
\par After removing the CPP, $(\ref{r[n]})$ can be written in matrix form as follows
\begin{equation}
  \mathbf{r}=\mathbf{Hs}+\mathbf{w}
\end{equation}
Taking into account that $\mathbf{w}$ follows a complex circularly-symmetric Gaussian distribution $\mathcal{C}\mathcal{N}\left(0, N_0\right)$, the channel matrix $\mathbf{H}$ is given by $\mathbf{H} = \sum_{i=1}^P{h_i\mathbf{\Gamma}_{\mathrm{CPP}_i}\mathbf{\Delta}_{f_i}\mathbf{\Pi}^{l_i}}$, where $\mathbf{\Pi}$ denotes the forward cyclic-shift matrix
\begin{equation}
  \mathbf{\Pi} =\left[ \begin{matrix}
    0&		\cdots&		0&		1\\
    1&		\cdots&		0&		0\\
    \vdots&		\ddots&		\vdots&		\vdots\\
    0&		\cdots&		1&		0\\
  \end{matrix} \right] _{N\times N}
\end{equation}
$\mathbf{\Delta} _{f_i}\triangleq \mathrm{diag}\left( e^{-j2\pi f_in},n=0,1,\dots ,N-1 \right) $ and $\mathbf{\Gamma} _{\mathrm{CPP}_i}$ is an $N\times N$ diagonal matrix
\begin{equation}
  \label{gamma}
  \begin{aligned}
    \mathbf{\Gamma} _{\mathrm{CPP}_i}=\mathrm{diag}( \begin{cases}
      e^{-j2\pi c_1\left( N^2-2N\left( l_i-n \right) \right)}&		,n<l_i\\
      1&		,n\ge l_i\\
    \end{cases}\\
    ,n=0,\dots ,N-1 )
  \end{aligned}
\end{equation}
As deduced from $(\ref{gamma})$, it becomes apparent that when $2Nc_1$ holds an integer value and $N$ is even, the matrix $\mathbf{\Gamma}_{\mathrm{CPP}_i}$ reduces to the identity matrix, i.e. $\mathbf{\Gamma}_{\mathrm{CPP}_i} = \mathbf{I}$.
\subsubsection{Demodulation}
After channel transmission, the signal received at the receiving end is demodulated to obtain the following
\begin{equation}
  \label{ym}
  y\left[ m \right] =\frac{1}{\sqrt{N}}\sum_{n=0}^{N-1}{e^{-j2\pi \left( c_1n^2+\frac{nm}{N}+c_2m^2 \right)}\cdot r\left[ n\right]}
\end{equation}
where $m=0,\dots ,N-1$. Writing $(\ref{ym})$ in the form of a matrix yields
\begin{equation}
  \label{y}
  \mathbf{y}=\mathbf{Ar}=\sum_{i=1}^P{h_i\mathbf{A}\mathbf{\Gamma }_{\mathrm{CPP}_i}\mathbf{\Delta} _{f_i}\mathbf{\Pi} ^{l_i}\mathbf{A}^H\mathbf{x}}+\mathbf{Aw}=\mathbf{H}_{\mathrm{eff}}\mathbf{x}+\tilde{\mathbf{w}}
\end{equation}
where $\mathbf{H}_{\mathrm{eff}}\triangleq \mathbf{AHA}^H$ and $\tilde{\mathbf{w}}=\mathbf{Aw}$. Since matrix $\mathbf{A}$ is unitary, the statistical characteristics of $\tilde{\mathbf{w}}$ remain unchanged and maintain the same statistical properties as $\mathbf{w}$.
\section{MESSAGE PASSING ALGORITHM based AFDM Communication}
In this section, we first provide the input-output relation of AFDM based on $(\ref{y})$, and then propose the MP algorithm based on this input-output relation.
\subsection{Input-Output Relation}
Referring to the input-output relation in $(\ref{y})$, it becomes evident that the received signal is formed as a linear combination of the transmitted signal. According to the definition of $\mathbf{H}_{\mathrm{eff}}$, we can write $(\ref{y})$ as follows
\begin{equation}
  \label{yi}
  \mathbf{y}=\sum_{i=1}^P{h_i\mathbf{H}_i\mathbf{x}}+\tilde{\mathbf{w}}
\end{equation}
where $\mathbf{H}_i\triangleq \mathbf{A}\mathbf{\Gamma} _{\mathrm{CPP}_i}\mathbf{\Delta} _{f_i}\mathbf{\Pi} ^{l_i}\mathbf{A}^H$. Evidently, it is observable that $H_i[p,q]$ can be represented as
\begin{equation}
  \label{Hi}
  H_i\left[ p,q \right] =\frac{1}{N}e^{j\frac{2\pi}{N}\left( Nc_1l_{i}^{2}-ql_i+Nc_2\left( q^2-p^2 \right) \right)}\mathcal{F} _i\left( p,q \right)
\end{equation}
where we define $\mathcal{F} _i\left( p,q \right) $ as follws due to $\nu_i$ is an integer
\begin{equation}
  \mathcal{F} _i\left( p,q \right) =\begin{cases}
    N&		,q=\left( p+\mathrm{loc}_i \right) _N\\
    0&		,\mathrm{otherwise}\\
  \end{cases}
\end{equation}
where $\mathrm{loc}_i\triangleq \left( \alpha _i+2Nc_1l_i \right) _N$, $\alpha_i$ is equivalent to $\nu_i$ within the context of this paper, the operation $\left( \cdot \right)_N $ corresponds to the modulo $N$ operation. Thus, $(\ref{Hi})$ can be expressed as follows
\begin{equation}
  H_i\left[ p,q \right] =\begin{cases}
    e^{j\frac{2\pi}{N}\left( Nc_1l_{i}^{2}-ql_i+Nc_2\left( q^2-p^2 \right) \right)}&	,q=\left( p+\mathrm{loc}_i \right) _N\\
    0&	,\mathrm{otherwise}\\
  \end{cases}
\end{equation}
From the above equation, it can be seen that each row of $\mathbf{H}_i$ has only one element that is nonzero as shown in Fig. 2. And we can write $(\ref{yi})$ as follows
\begin{equation}
  \label{yp}
  \begin{aligned}
    y\left[ p \right] =\sum_{i=1}^P{h_ie^{j\frac{2\pi}{N}\left( Nc_1l_{i}^{2}-ql_i+Nc_2\left( q^2-p^2 \right) \right)}x\left[ q \right] +\tilde{w}\left[ p \right]}\\
    , 0\le p\le N-1
  \end{aligned}
\end{equation}
where $q=\left( p+\mathrm{loc}_i \right) _N$.
\begin{figure}[h!]
  \centering
  \includegraphics[width=2in]{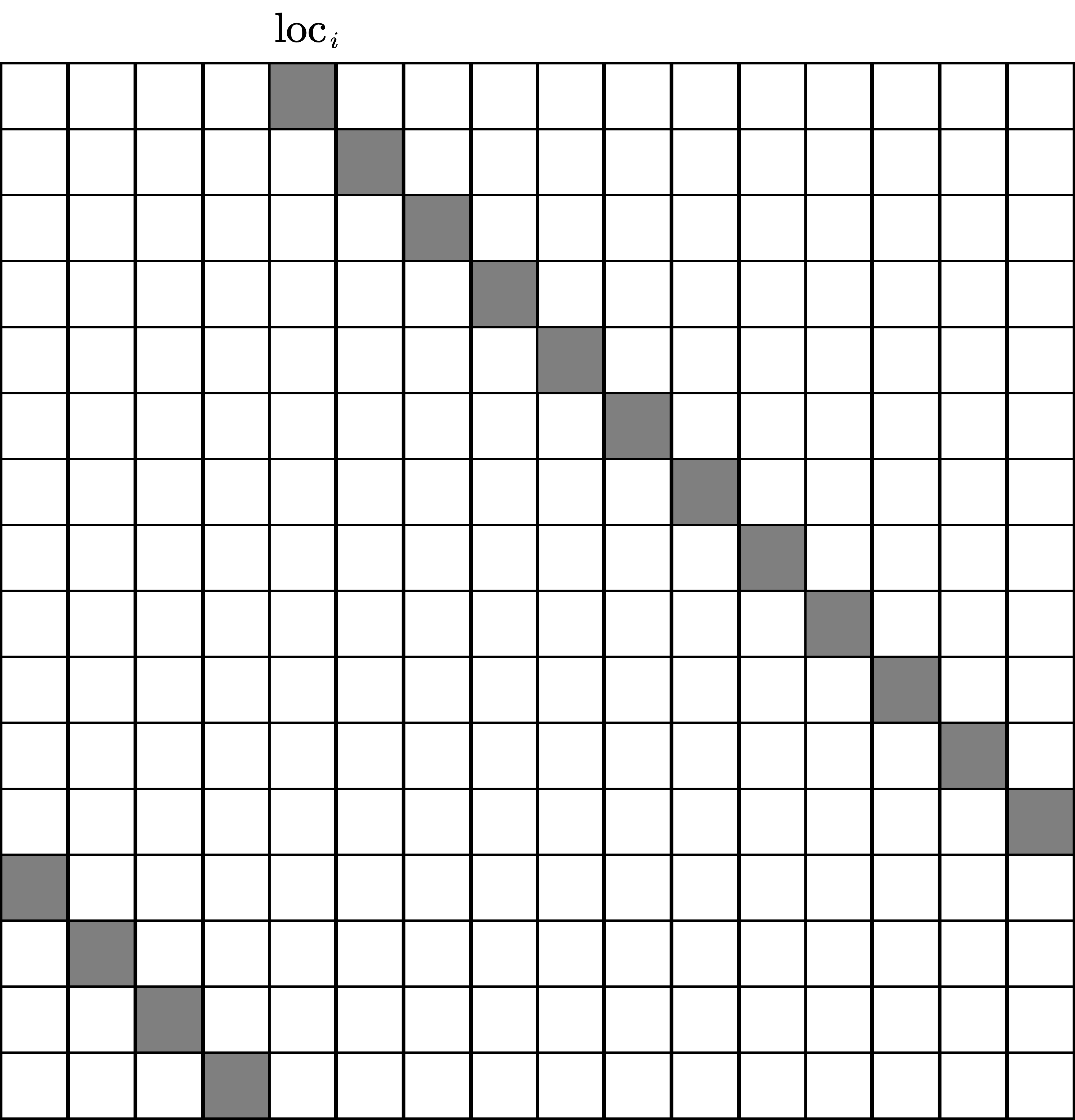}
  \caption{Structure of $\mathbf{H}_i$ for integer Doppler shifts.}
\end{figure}
\subsection{MP based Detection}
According to the input-output relation of vectorized AFDM in $(\ref{y})$, it can be seen that $\mathbf{y}$ and $\tilde{\mathrm{w}}$ are complex vectors with $N$ elements, represented by $y[d]$ and $\tilde{w}\left[ d \right] $, respectively, where $1\le d\le N$; $\mathbf{H}_{\mathrm{eff}}$ is an $N$-order complex matrix, with elements represented by $H_{\mathrm{eff}}[d,c]$, where $1\le d,c\le N$; $\mathbf{x}$ is an information vector with $N$ elements, represented by $x\left[ c \right]$, where $x\left[ c \right] \in \mathbb{A}$, $1\le c\le N$. The values of the element of $\mathbf{y}$, $\mathbf{x}$ and $\mathbf{H}_{\mathrm{eff}}$ can be determined by $(\ref{yp})$, and $\tilde{\mathrm{w}}$ is the noise vector. According to $(\ref{yp})$, we observe that among the $N$ elements in row $d$ of $\mathbf{H}_{\mathrm{eff}}$, it is non-zero in $(d+\mathrm{loc}_i)_N$, and among the $N$ elements in column $c$, it is non-zero in $(c-\mathrm{loc}_i)_N$, where $i$ denotes the $i$-th path. We define the index set of non-zero elements in row $d$ of $\mathbf{H}_{\mathrm{eff}}$ as $I(d)$ and the index set of non-zero elements in column $c$ of $\mathbf{H}_{\mathrm{eff}}$ as $J(c)$.
\par According to $(\ref{y})$, the system model can be regarded as a factor graph with sparse connections. Here, the vector $\mathbf{x}$ has $N$ variable nodes, while vector $\mathbf{y}$ consists of $N$ observation nodes. In this factor graph, we can see that each observation node $y[d]$ maintains a set of $P$ connected variable nodes denoted by $\{x[c],c \in I(d)\}$. Similarly, every variable node $x[c]$ establishes connections with a set of $P$ associated observation nodes designated as $\{y[d],d \in J(c)\}$. The parameter $P\ge 1$ signifies the number of paths in this context.
\par From $(\ref{y})$, we can derive a detection rule for estimating the joint maximum a posterior probability (MAP) as follows
$$
\hat{\mathbf{x}}=\underset{\mathbf{x}\in \mathbb{A} ^{N\times 1}}{\mathrm{arg\max}}\mathrm{Pr}\left( \mathbf{x}\left| \mathbf{y},\mathbf{H}_{\mathrm{eff}} \right. \right)
$$
The complexity of this approach grows exponentially with $N$. Due to the fact that joint MAP detection may be difficult to handle with actual $N$ values, we instead contemplate a symbol-by-symbol MAP detection strategy for $c=1,\dots,N$
\begin{subequations}
  \begin{align}
    \hat{x}\left[ c \right] &=\mathrm{arg} \underset{a_j\in \mathbb{A}}{\max}\mathrm{Pr}\left( x\left[ c \right] =a_j|\mathbf{y},\mathbf{H}_{\mathrm{eff}} \right)  \notag\\
    &=\mathrm{arg} \underset{a_j\in \mathbb{A}}{\max}\frac{1}{Q}\mathrm{Pr}\left( \mathbf{y}|x\left[ c \right] =a_j,\mathbf{H}_{\mathrm{eff}} \right)  \label{xca}\\
    &\approx \mathrm{arg} \underset{a_j\in \mathbb{A}}{\max}\prod_{d\in J_c}{\mathrm{Pr}\left( y\left[ d \right] |x\left[ c \right] =a_j,\mathbf{H}_{\mathrm{eff}} \right)} \label{xcb}
  \end{align}
\end{subequations}
Due to the sparsity of $\mathbf{H}_{\mathrm{eff}}$, in $(\ref{xca})$, we assume equal probabilities for all transmitted symbols $a_j \in \mathbb{A}$. Furthermore, in $(\ref{xcb})$, we assume that for a given $x[c]$, the components of vector $\mathbf{y}$ exhibit a degree of approximate independence. That is, we make the assumption that the interference terms, denoted as $\zeta^{(i)}_{d,c}$ and defined in $(\ref{zeta})$, are independent for a specific value of $c$.
\par To address the approximation challenge associated with symbol-by-symbol MAP detection as described in $(\ref{xcb})$, we introduce an MP detector exhibiting linear complexity concerning $N$. For each observed $y[d]$, the variable $x[c]$ is separated from the remaining interference terms and treated as Gaussian noise. This noise is characterized by a mean and variance that can be computed with ease.
\begin{figure}[h!]
  \centering
  \includegraphics[width=3in]{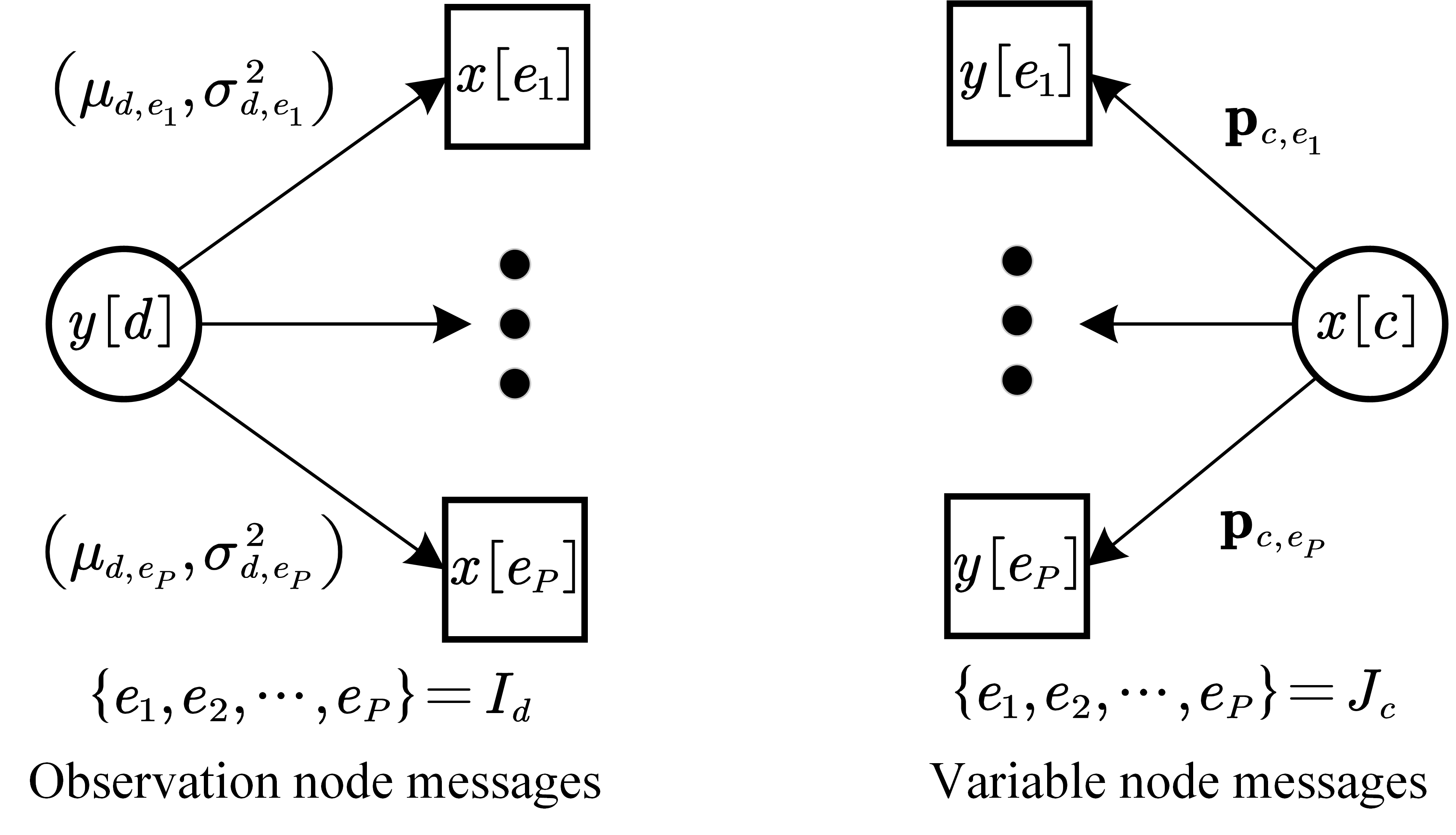}
  \caption{Factor graph of AFDM system.}
\end{figure}
\par Within the framework of the MP algorithm, the information transmitted from observation nodes to variable nodes involves the mean and variance of the interference terms. Conversely, the information conveyed from a variable node $x[c]$ and directed towards observation nodes $y[d]$, where $d \in J(c)$, encompasses the probability mass function (pmf) of the alphabet, denoted as $\mathbf{p}_{c,d} = \{p_{c,d}(a_j)|a_j \in \mathbb{A}\}$. The graphical representation in Fig. 3 illustrates the connections and the information exchanges taking place between the observation and variable nodes. The detailed steps of the MP algorithm are outlined in Algorithm 1 provided below.
\begin{algorithm}
  \caption{MP Algorithm for AFDM Symbol Detection}
  \begin{algorithmic}[1]
  \renewcommand{\algorithmicrequire}{\textbf{Inputs:}}
  \renewcommand{\algorithmicensure}{\textbf{Output:}}
  \REQUIRE Channel matrix $\mathbf{H}_{\mathrm{eff}}$, received signal $\mathbf{y}$, the maximum iterations $i_{\max}$
  \ENSURE Estimated signal $\hat{x}[c]$.
  \STATE \textbf{Initialize:} $\mathrm{pmf}$ $\mathbf{p}^{(0)}_{c,d}=\frac{1}{Q}$, $c=0,\dots,N-1$, $d\in J(c)$
  \FOR {$i=1:i_{\max}$}
  \STATE Observation nodes $y[d]$ uses $\mathbf{p}^{(i-1)}_{c,d}$ to calculate the means $\mu_{d,c}^{(i)}$ and variances $(\sigma^{(i)}_{d,c})^2$ of the Gaussian random variables $\zeta^{(i)}_{d,c}$ and passes them to variable nodes $x[c]$, $c \in I(d)$.
  \STATE Variable nodes $x[c]$ revise $\mathbf{p}^{(i)}_{c,d}$ using $\mu_{d,c}^{(i)}$, $(\sigma^{(i)}_{d,c})^2$ and $\mathbf{p}^{(i-1)}_{c,d}$ and passes them to observation nodes $y[d]$, $d \in J(c)$.
  \STATE Calculate convergence indicator $\eta^{(i)}$.
  \STATE Revise the decision on $\hat{x}[c]$.
  \IF {(Stopping criteria satisfied)}
  \STATE \textbf{EXIT}
  \ENDIF
  \ENDFOR
  \end{algorithmic}
\end{algorithm}\\
The specific steps for the $i$-th iteration of the MP algorithm are as follows
\vspace{1mm}
\par \textbf{1. Transmiting from observation nodes $y[d]$ to variable nodes $x[c]$, $c \in I(d)$:}
\par The Gaussian random variable $\zeta^{(i)}_{d,c}$ is defined as follows
\begin{equation}
  \label{zeta}
  \zeta _{d,c}^{(i)}=\sum_{e\in I\left( d \right) ,e\ne c}{x\left[ e \right] H_{\mathrm{eff}}\left[ d,e \right]}+\tilde{w}\left[ d \right]
\end{equation}
From $(\ref{zeta})$, the mean $\mu_{d,c}^{(i)}$ and variance $(\sigma^{(i)}_{d,c})^2$ of the interference can be calculated as
\begin{equation}
  \mu _{d,c}^{(i)}=\sum_{e\in I\left( d \right) ,e\ne c}{\sum_{j=1}^Q{p_{e,d}^{(i-1)}\left( a_j \right) a_j H_{\mathrm{eff}}\left[ d,e \right]}}
\end{equation}
and
\begin{equation}
  \begin{aligned}
    \left( \sigma _{d,c}^{(i)} \right) ^2=\sum_{e\in I\left( d \right) ,e\ne c}{\left( \sum_{j=1}^Q{p_{e,d}^{(i-1)}\left( a_j \right) \left| a_j \right|^2\left| H_{\mathrm{eff}}\left[ d,e \right] \right|^2} \right.} \\
    \left. -\left| \sum_{j=1}^Q{p_{e,d}^{(i-1)}\left( a_j \right) a_j H_{\mathrm{eff}}\left[ d,e \right]} \right|^2 \right) +\sigma ^2
  \end{aligned}
\end{equation}
From $(\ref{yp})$, we can calculate $H_{\mathrm{eff}}[d,e]$ as follows
\begin{equation}
  H_{\mathrm{eff}}[d,e] = h_ie^{j\frac{2\pi}{N}\left(Nc_1l^2_i-ql_i+Nc_2\left(q^2-d^2\right)\right)}
\end{equation}
where $q=(d+\mathrm{loc}_i)_N$ and $i$ is the $i$-th path corresponding to $e$.
\vspace{1mm}
\par \textbf{2. The transmission of information from variable nodes $x[c]$ to observation nodes $y[d]$, $d\in J(c)$:}
\par We can revise the pmf vector $\mathbf{p}^{(i)}_{c,d}$ as follows
\begin{equation}
  p_{c,d}^{(i)}\left( a_j \right) =\Delta \cdot \tilde{p}_{c,d}^{(i)}\left( a_j \right) +\left( 1-\Delta \right) \cdot p_{c,d}^{(i-1)}\left( a_j \right) ,a_j\in \mathbb{A}
\end{equation}
where the damping factor $\Delta \in (0, 1]$\cite{delta} is employed to enhance performance by regulating the speed of convergence, and
\begin{equation}
  \begin{aligned}
    \tilde{p}_{c,d}^{(i)}\left( a_j \right) &\propto \prod_{e\in J\left( c \right) ,e\ne d}{\mathrm{Pr}\left( y\left[ e \right] |x\left[ c \right] =a_j,\mathbf{H}_{\mathrm{eff}} \right)} \\
  &=\prod_{e\in J\left( c \right) ,e\ne d}{\frac{\xi ^{(i)}\left( e,c,j \right)}{\sum_{k=1}^Q{\xi ^{(i)}\left( e,c,k \right)}}}
  \end{aligned}
\end{equation}
where $\xi ^{(i)}\left( e,c,k \right) =\exp \left( \frac{-\left| y\left[ e \right] -\mu _{e,c}^{(i)}-H_{\mathrm{eff}}\left[ e,c \right] a_k \right|^2}{\left( \sigma _{e,c}^{(i)} \right) ^2} \right) $. From $(\ref{yp})$, we can calculate $H_{\mathrm{eff}}[e,c]$ as follows
\begin{equation}
  H_{\mathrm{eff}}[e,c] = h_ie^{j\frac{2\pi}{N}\left(Nc_1l^2_i-cl_i+Nc_2\left(c^2-p^2\right)\right)}
\end{equation}
where $p=(c-\mathrm{loc}_i)_N$ and $i$ is the $i$-th path corresponding to $e$.
\vspace{1mm}
\par \textbf{3. Calculate convergence indicator: }We can calculate the convergence indicator $\eta^{(i)}$ as follows
\begin{equation}
  \eta ^{(i)}=\frac{1}{N}\sum_{c=1}^N{\mathbb{I} \left( \underset{a_j\in \mathbb{A}}{\max}\; p_{c}^{(i)}\left( a_j \right) \ge 1-\gamma \right)}
\end{equation}
for a certain small $\gamma>0$ and where
\begin{equation}
  p_{c}^{(i)}\left( a_j \right) =\prod_{e\in J\left( c \right)}{\frac{\xi ^{(i)}\left( e,c,j \right)}{\sum_{k=1}^Q{\xi ^{(i)}\left( e,c,k \right)}}}
\end{equation}
and the symbol $\mathbb{I} \left( \cdot \right)$ denotes an indicator function. The indicator function outputs 1 when the expression inside is true and 0 when it is false.
\vspace{1mm}
\par \textbf{4. Revise the decision on $\hat{x}[c]$: }When $\eta^{(i)} > \eta ^{(i-1)}$, the decision concerning the transmitted symbol is adjusted as follows
\begin{equation}
  \hat{x}\left[ c \right] =\mathrm{arg} \underset{a_j\in \mathbb{A}}{\max}\;p_{c}^{(i)}\left( a_j \right) , c=0,\cdots ,N-1
\end{equation}
We revise the decision about transmitted symbols solely when the ongoing iteration is capable of offering superior estimates compared to the preceding iteration.
\vspace{1mm}
\par \textbf{5. Stopping conditions: }The MP algorithm concludes its execution when any of the subsequent conditions are satisfied:
\par \textbf{(\romannumeral1)}\quad $\eta^{(i)}=1$,
\par \textbf{(\romannumeral2)}\quad $\eta ^{(i)}<\eta ^{(i^*)}-\epsilon $, where $i^*\in \{1,\dots,i-1\}$ is the iteration index corresponding to the maximum value of $\eta ^{(i^*)}$,
\par \textbf{(\romannumeral3)}\quad The algorithm has reached the maximum number of iterations $i_{\max}$.
\par The value $\epsilon = 0.2$ is chosen to disregard minor fluctuations in $\eta$. In the best-case scenario, the first condition is fulfilled when all symbols have reached convergence. When the ongoing iteration results in a less favorable decision compared to previous iterations, the \textbf{(\romannumeral2)} condition is intended to stop the algorithm.
\section{SIMULATION RESULTS}
In this section, we conducted extensive simulations and assessed the performance of AFDM by analyzing the obtained simulation results. Throughout all simulations, the parameters $c_1$ and $c_2$ within the context of DAFT are assigned values of $\left( 2\alpha _{\max}+1 \right) /N$ and 0, respectively, where $\alpha_{\max}$ represents the upper limit of Doppler shift within the channel, while $N$ signifies the number of symbols in AFDM. The complex gain $h_i$ is generated as an independent complex Gaussian random variable with a mean of zero and a variance of $1/P$, where $P$ is a constant. To obtain the bit error rate (BER) value, $10^6$ distinct channel realizations are employed.
\begin{figure}[!t]
  \centering
  \vspace*{-3mm}
  \subfloat[The variation of BER with $\Delta$.]{
		\includegraphics[width=3in]{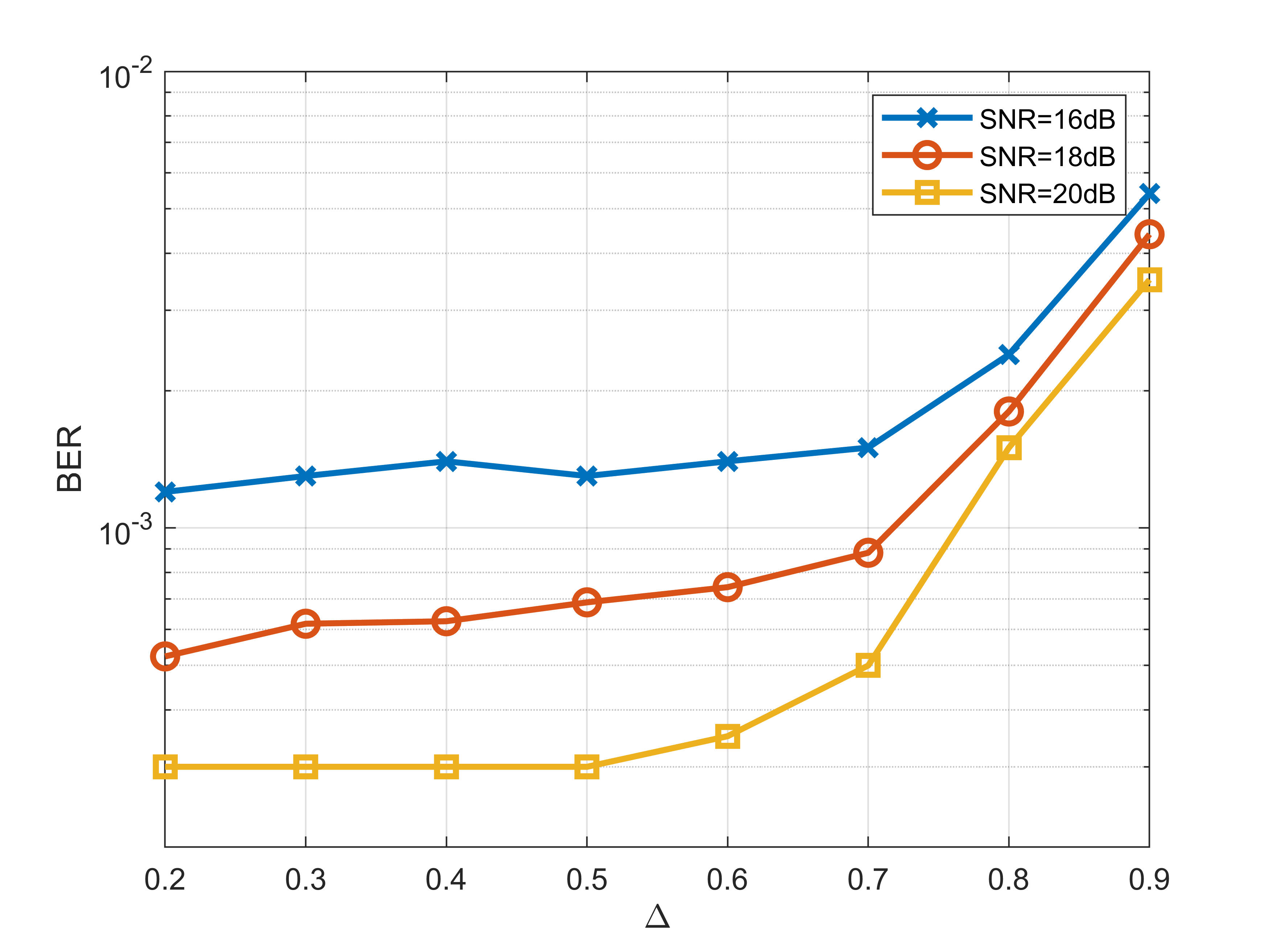}}\\
    \vspace*{-3mm}
  \subfloat[The variation of average number of iterations with $\Delta$.]{
		\includegraphics[width=3in]{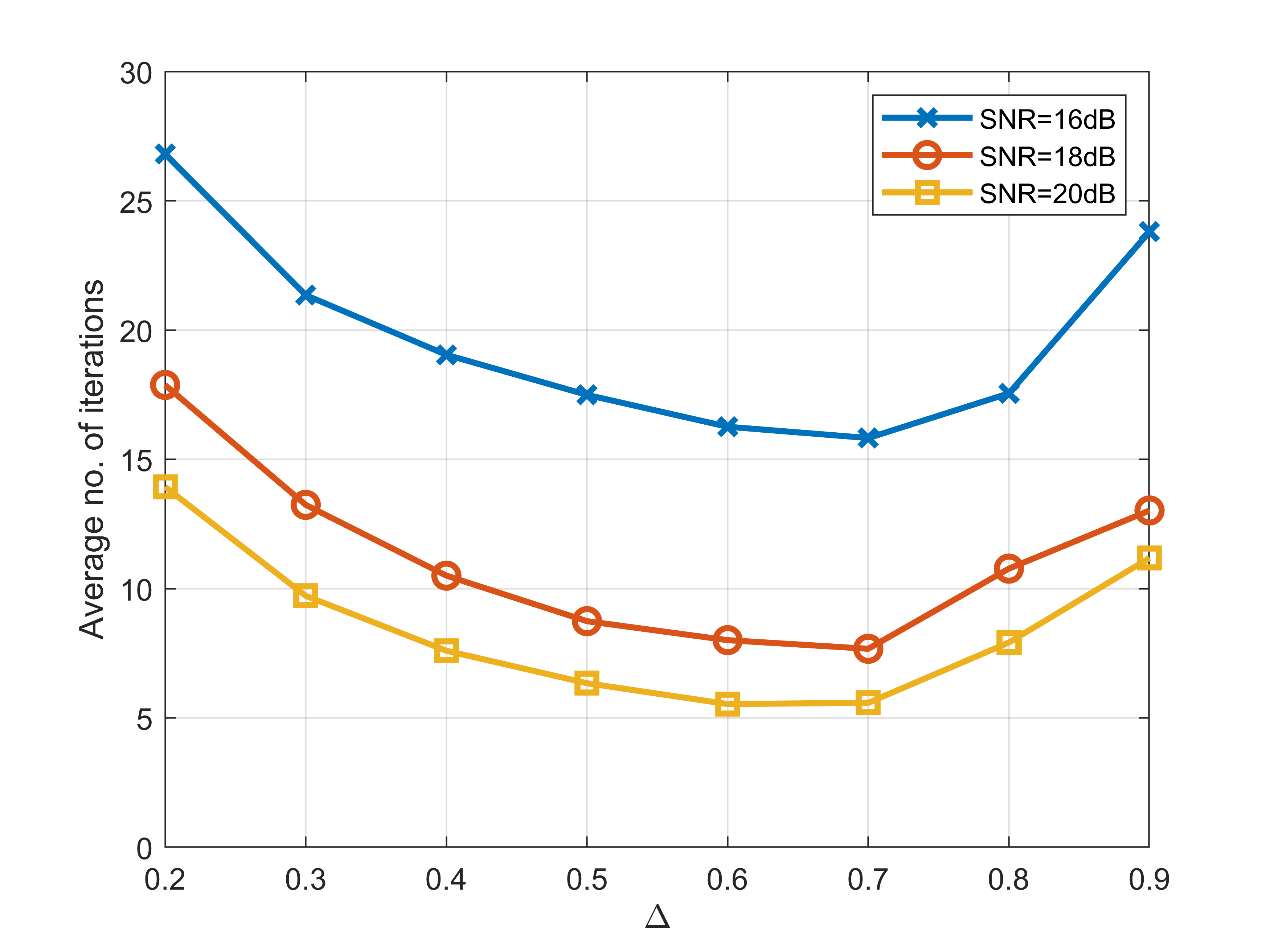}}
  \caption{The variation of BER and average number of iterations with $\Delta$.}
  \vspace*{-4mm}
\end{figure}
\par Fig. 4 depicts variations of BER performance and the average number of iterations for different values of the damping factor $\Delta$ while using the MP algorithm with ideal pulses. The analysis involves varying the damping factor $\Delta$ while considering the 4-QAM signal. The SNR is set to 16 dB, 18 dB, and 20 dB, respectively. From Fig. 4(a), it is evident that the BER remains relatively stable when $\Delta\le 0.6$, but it starts to deteriorate beyond this value. Conversely, as depicted in Fig. 4(b), it is evident that the MP algorithm attains convergence with the minimum number of iterations when $\Delta=0.7$. Drawing from these findings, we deduce that the optimal damping factor, considering both performance and complexity, lies at $\Delta=0.6$. Consequently, we opt for this value in our AFDM system.
\begin{figure}[h!]
  \centering
  \includegraphics[width=3in]{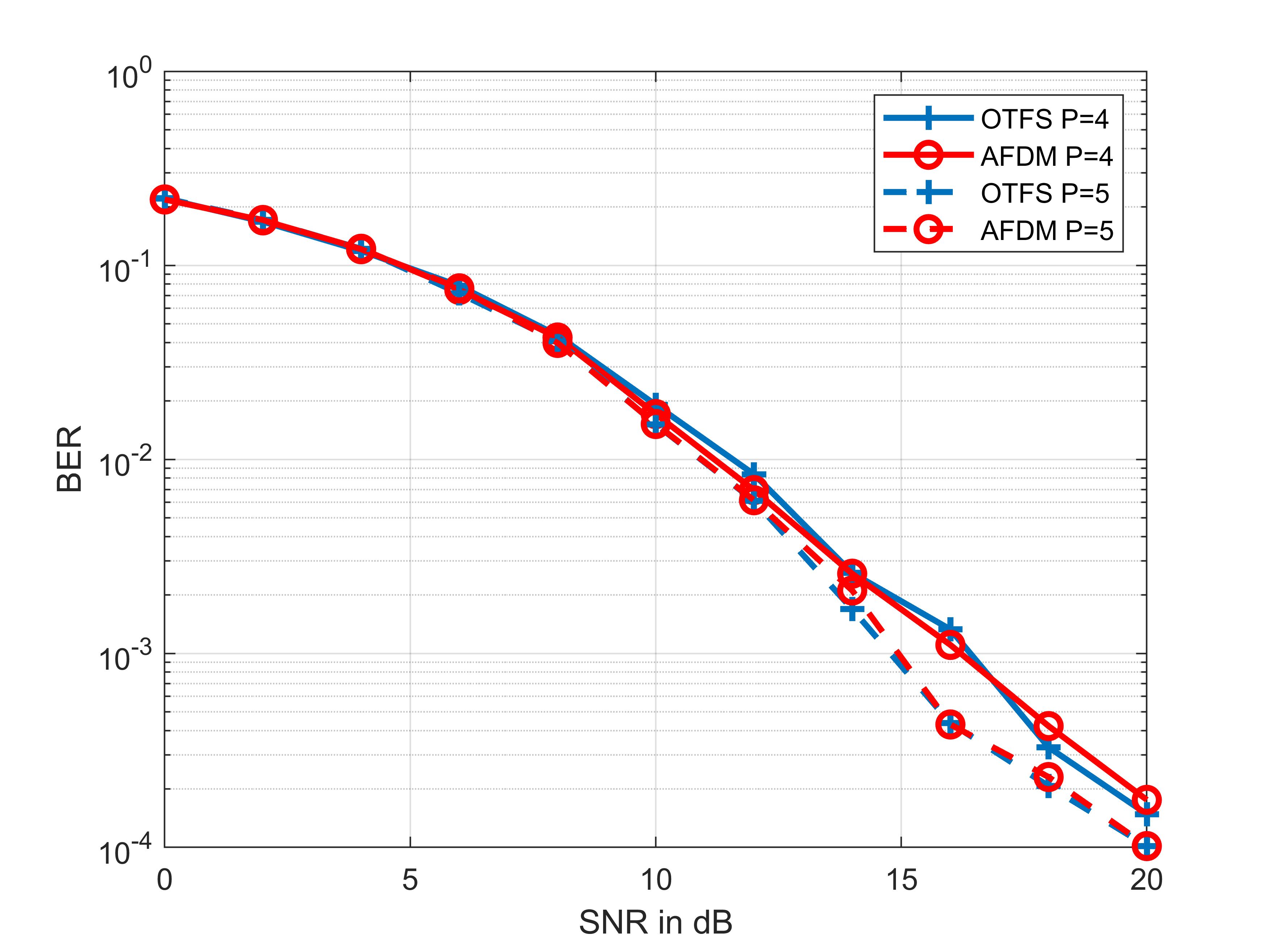}
  \caption{The BER performance of 4-QAM AFDM with $N=64$ and 4-QAM OTFS with $N_{\rm{OTFS}}=8$, $M_{\rm{OTFS}}=16$ using MP detection in 4-path LTV channels and 5-path LTV channels with $l_{\max}=\alpha_{\max}=3$.}
\end{figure}
\par Fig. 5 depicts two schemes for AFDM with $N=128$ and OTFS with $N_{\rm{OTFS}}=8$ and $M_{\rm{OTFS}}=16$, both utilizing 4-QAM symbols and MP detection, to assess the BER performance in 4-path and 5-path LTV channels, both with $l_{\max}=\alpha_{\max}=3$. From the results, it is evident that as the number of paths increases, the amount of information during MP detection also increases. Consequently, the BER performance is better at $P=5$ than at $P=4$. Additionally, the results show that the BER performance of AFDM and OTFS remains similar, regardless of whether $P=4$ or $P=5$ is considered. This similarity in BER performance showcases the potential of AFDM as a competitive alternative to OTFS in scenarios with varying path conditions.
\begin{figure}[h!]
  \centering
  \vspace*{-3mm}
  \includegraphics[width=3in]{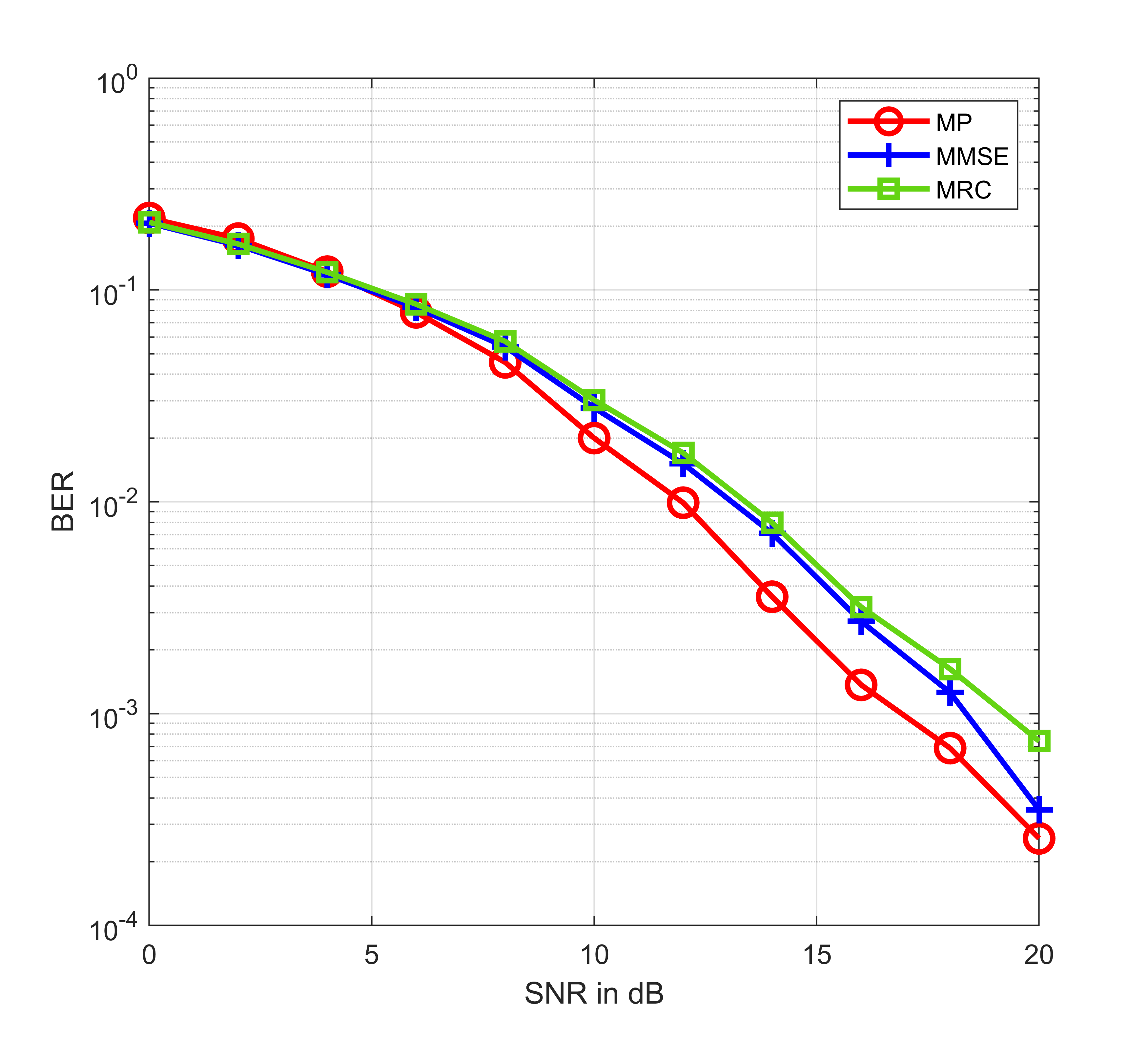}
  \caption{The BER performance of 4-QAM AFDM with $N=64$ using MP, MMSE, and MRC detection in 4-path LTV channels with $l_{\max}=3$ and $\alpha_{\max}=3$.}
  \vspace*{-4mm}
\end{figure}
\par Fig. 6 demonstrates AFDM with $N=64$, employing 4-QAM signals, and utilizing MP detection, MMSE detection, and MRC detection to evaluate the BER performance in 4-path LTV channels with $l_{\max}=3$ and $\alpha_{\max}=3$. From the results, we have observed that the performance of MRC is relatively comparable to that of MMSE, indicating that MRC provides a decent BER performance. However, the MP detection outperforms both MMSE and MRC, demonstrating its superiority in achieving lower BER values compared to the other two detection algorithms.
\begin{figure}[h!]
  \centering
  \vspace*{-3mm}
  \includegraphics[width=3in]{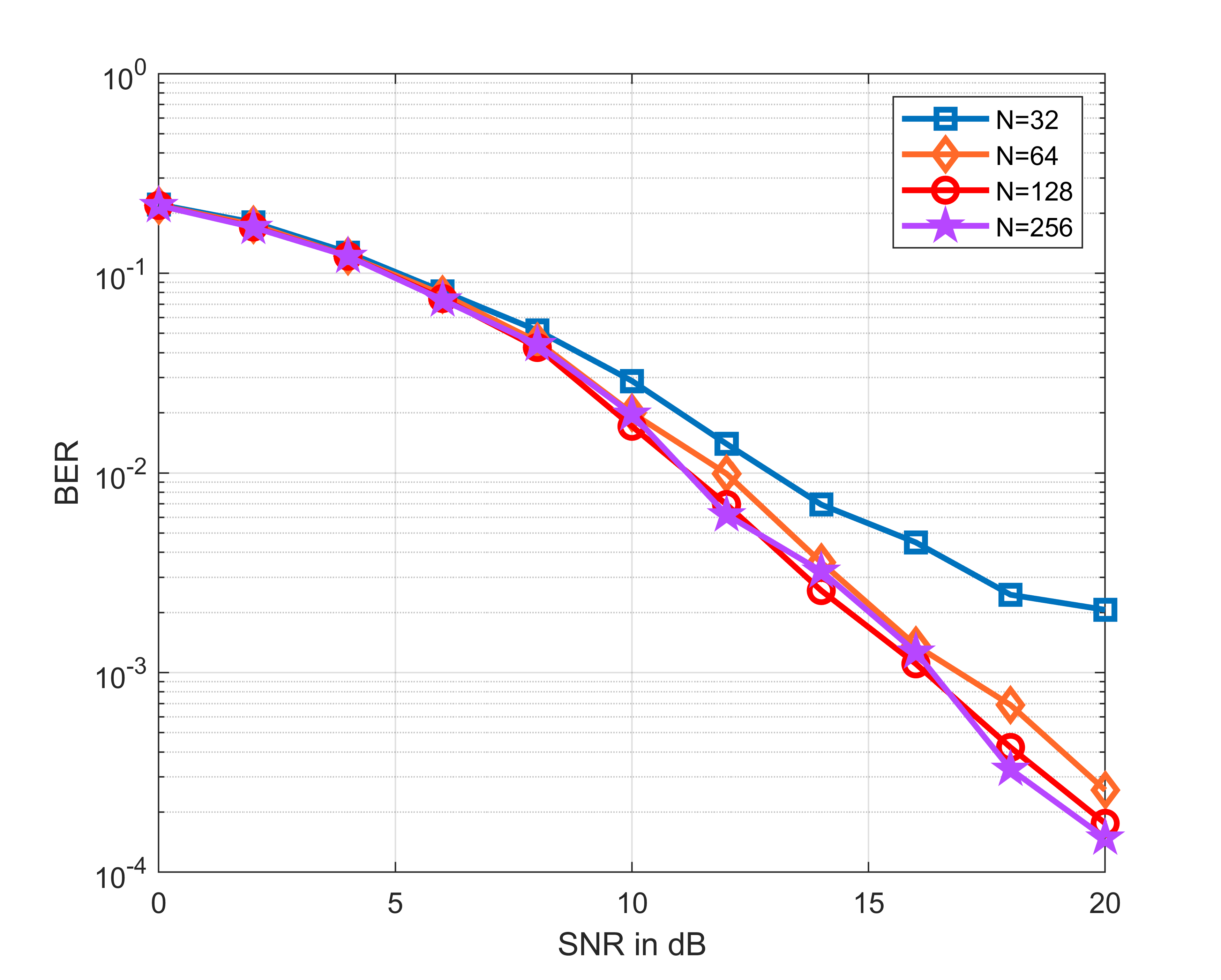}
  \caption{The BER performance of 4-QAM AFDM with $N=32$, $N=64$, $N=128$, and $N=256$ using MP detection in 4-path LTV channel with $l_{\max}=3$ and $\alpha_{\max}=3$.}
\end{figure}
\par Fig. 7 illustrates the BER performance of AFDM with various numbers of symbol using MP detection in 4-path LTV channels with $l_{\max}=3$ and $\alpha_{\max}=3$. The considered AFDM systems have different symbol numbers, namely $N=32$, $N=64$, $N=128$, and $N=256$. From the observations in Fig. 7, we can conclude that as the symbol number $N$ increases, the BER of the AFDM system decreases, and the MP detection effect improves. In other words, larger $N$ values result in better BER performance and more robust detection for the AFDM system using MP detection.

\section{Conclusion}
In this paper, we introduce an innovative MP algorithm that is both low-complexity and efficient, specifically designed for joint symbol detection in large-scale AFDM systems. The MP algorithm effectively addresses challenges such as ISI and ICI through suitable phase adjustments, while also mitigating IDI by focusing on significant interference terms. Furthermore, the proposed MP algorithm exhibits remarkable compensation capabilities for wide-ranging channel Doppler spread. Through simulation results, we demonstrate that the BER performance of AFDM with MP detection closely matches that of OTFS modulation, while outperforming MMSE detection. Furthermore, our observations indicate that AFDM's BER performance improves with larger symbol numbers $N$ and enhanced MP detection efficiency. These findings underscore the potential of our proposed MP algorithm in elevating the performance of AFDM systems.

\end{document}